\documentstyle{article}
\def\beq{\begin{equation}}
\def\eeq{\end{equation}}
\def\bea{\begin{eqnarray}}
\def\eea{\end{eqnarray}}
\def\nb{{\overline N}}
\begin{document}
\begin{titlepage}
\begin{flushright}   LBNL-39745 \\ UCB-PTH-96/62 \\ UFIFT-HEP-96-32 \\ 
LPTHE-ORSAY 96/101 \\ hep-ph/9612442 \end{flushright}
\vskip .8cm
\centerline{\LARGE{\bf {Anomalous $U(1)$ and low-energy physics:}}}
\vskip .5cm
\centerline{\LARGE{\bf {the power of D-flatness and holomorphy}}}
\vskip 1.5cm
\centerline{\bf {Pierre Bin\'etruy${}^{a,b)}$\footnote{Visiting Miller
professor at the Department of Physics of the University of California,
Berkeley}, Nikolaos Irges${}^{c)}$, St\'ephane Lavignac${}^{a)}$\footnote{Supported in
part by the Human Capital and Mobility
Programme, contract CHRX-CT93-0132.} 
and Pierre Ramond${}^{c)}$\footnote{Supported in part by the
United States Department of Energy under grant DE-FG05-86-ER40272.}}}   \vskip .5cm
\centerline{\em ${}^{a)}$ Laboratoire de Physique Th\'eorique et Hautes
Energies\footnote{Laboratoire associ\'e au CNRS-URA-D0063.}}
\centerline{\em Universit\'e Paris-Sud, B\^at. 210,}
\centerline{\em F-91405 Orsay Cedex, France }
\vskip .5cm
\centerline{\em ${}^{b)}$ Department of Physics}
\centerline{\em and Theoretical Physics Group, Lawrence Berkeley National Laboratory,}
\centerline{\em University of California,}
\centerline{\em Berkeley CA 94720, USA}
\vskip .5cm
\centerline{\em ${}^{c)}$ Institute for Fundamental Theory,}
\centerline{\em Department of Physics, University of Florida}
\centerline{\em Gainesville FL 32611, USA}

\vskip 2cm
\centerline{\bf {Abstract}}
\vskip 2cm

In models with an anomalous abelian symmetry broken at a very large 
scale, we study which requirements to impose on the anomalous charges in 
order to prevent standard model fields from  acquiring large vacuum 
expectation values. The use of holomorphic invariants to study $D$-flat 
directions for the anomalous symmetry, proves to be a very powerful tool. 
We find that in order to forbid unphysical vacuum configurations at that 
scale, the superpotential must contain many interaction terms, including 
the usual Yukawa terms.   Our analysis suggests that the 
anomalous charge of the $\mu$-term is zero. It is remarkable that, 
together with the seesaw mechanism, and mass hierarchies, this implies a 
natural conservation of $R$-parity. 
 \indent 
\end{titlepage}


\section{Introduction}
Many effective superstring models have an  anomalous  Abelian 
gauge symmetry.  Its anomalies are cancelled through  the 
four-dimensional field theory remnant~\cite{DSW} of the Green-Schwarz 
mechanism~\cite{GS}, and it is naturally broken at a scale $\xi$, a small 
computable factor times the Planck scale.

This anomalous symmetry may play a role in various domains of relevance 
for the low energy world~\cite{Ib}-\cite{PMR}. For example, if one 
considers an Abelian family symmetry~\cite{FN,LNS} to try and explain the 
mass hierarchies observed in the quark and charged lepton sector, this 
symmetry is most probably anomalous~\cite{Bijnens,BR}. By naturally 
providing a small expansion parameter, $\xi/M_{Pl}$, this symmetry is 
suited to the analysis of lepton and quark mass hierarchies~\cite{RRR}. 
The same  symmetry might also  trigger  supersymmetry 
breaking~\cite{BD,DP,MR}. Its role has also been advocated in predicting 
neutrino mixing patterns~\cite{DLLRS,GN,BLPR}, and as possible solutions 
to the doublet-triplet splitting in grand unified models~\cite{DPok,BT}.  

In this context, it is especially important to characterize the flat 
directions along which the breaking of this anomalous symmetry occurs. In 
this letter, we wish to show that there are simple and powerful methods 
to undertake this task. Since $\xi$, the scale of breaking  of this 
$U(1)_X$ symmetry is close to the Planck scale, there are severe 
constraints from phenomenology: Supersymmetry  and the 
gauge symmetries of the standard model must remain unbroken. This means 
that all $D$ and $F$ terms must vanish, and that  no field of the 
standard model can get a vacuum expectation value (vev)  along these 
directions, since $\xi$ is much larger than the electroweak scale. This 
remains true even if the anomalous symmetry plays a role  in supersymmetry 
breaking. Indeed, in the work of Ref.~\cite{BD} where such a role is 
emphasized, it is found that $<D_X> \ll \xi^2$ which indicates that the 
analysis which follows is perfectly relevant to this case.

In a classic paper~\cite{invariant}, the correspondance between $D$-flat 
directions ({\em i.e.} field configurations for which $D$-terms vanish) 
and extrema of holomorphic invariant polynomials was 
established. These ideas were recently applied to the anomalous 
$U(1)$  by Dudas {\it et al}~\cite{DGPS}. Following these authors, we 
also  base our analysis on the construction of holomorphic invariants: 
invariants of the standard model~\cite{GKM}, and then, in order to 
discuss $D_X$-flatness (and $F$-flatness), invariants under the anomalous 
$U(1)_X$ symmetry.

The requirement that no standard model fields charged under the anomalous 
symmetry acquire vev's of order $\xi$  give powerful constraints on their 
anomalous charges, which in 
turn yield valuable information on several outstanding problems of low energy 
supersymmetry: the mu-problem, the origin of $R$-parity~\cite{Rparity} or 
the order of magnitude of its violations (and more generally of baryon 
and lepton number), and the neutrino mass generation through the seesaw 
mechanism~\cite{seesaw}. About the latter, we will see that right-handed 
neutrino superfields (in our discussion those are standard model singlets 
which do not get a vev at the scale $\xi$)  play an important role in the 
discussion of the $U(1)_X$-breaking directions.

We will start in the next Section by describing the method used and 
illustrating it on some examples, with  only one anomalous $U(1)_X$  broken 
at the scale $\xi$.  We then  turn in Section 3 to a discussion of the 
possible solutions for the mu-term. In Section 4, we show how these 
considerations naturally suggest  $R$-parity conservation. Finally in 
Section 5, we discuss the relevance of this to more complicated 
situations, such as complete models of mass hierarchies.

\section{The power of D-flatness and holomorphy}
We study  in what follows the directions along which occurs the breaking of an
anomalous $U(1)_X$ symmetry. The $U(1)_X$ D-term $D_X$ is of the form:
\begin{equation}
D_X \sim \sum_i X_i |\Phi_i|^2 - \xi^2
\label{eq:DX} \end{equation}
where $X_i$ is the $X$-charge of a generic scalar field $\Phi_i$ and $\xi$ is
the anomalous Fayet-Iliopoulos term. We consider three types of fields:
\begin{itemize}
\item fields which acquire a vacuum expectation value of order $\xi$ 
when the $U(1)_X$ symmetry
is broken: we denote them generically by $\theta$. 
\item fields charged under $SU(3) \times SU(2) \times U(1)$, typically the
fields found in the MSSM; these fields should not acquire vacuum expectation
values. They appear in invariants which form the building blocks used to construct
terms in the superpotential. Typically for the superfields in the MSSM:
\begin{eqnarray}
&H_d H_u\ ,&  {\bf Q}_i  \overline{\bf d}_kH_d\ , L_i H_d \bar e_k\ , {\bf Q}_i  
\overline{\bf u}_kH_u\ ,  \nonumber \\
&L_i H_u\ ,&  {\bf Q}_i  \overline{\bf d}_kL_j\ ,  L_i L_j \bar e_k\ ,  
\overline{\bf u}_i \overline{\bf d}_j \overline{\bf d}_k\ , 
\end{eqnarray}
where $i,j,k$ are family indices. We do not list the higher oder 
invariants which can be found in the literature~\cite{GKM}.  
We will denote these invariants generically by $S$.
\item scalar fields, singlet under the standard model gauge group, 
which do not receive
vacuum expectation values of order $\xi$. These fields are natural
candidates for the right-handed neutrinos and we will denote them by $\bar N$.
Typically,  for these fields to be interpreted as right-handed neutrinos, one 
needs terms in the superpotential to generate Majorana mass terms:
\begin{equation}
W_M \sim M \bar N^2 \left( {\theta / M}\right)^n
\label{eq:Majorana} \end{equation}
and terms to generate Dirac mass terms for the neutrino:
\begin{equation}
W_D \sim S \bar N \left( {\theta / M}\right)^p
\label {eq:Dirac} \end{equation}
where $S$ is the invariant $S= L H_u$. The presence of both terms 
(\ref{eq:Majorana}) and (\ref{eq:Dirac}) is necessary to implement the
seesaw mechanism~\cite{seesaw}. 
\end{itemize}
Finally, we will denote by $<\Phi_1, \Phi_2,\cdots,\Phi_n>$ the direction 
in scalar field parameter space where the fields $\Phi_1, 
\Phi_2,\cdots,\Phi_n$ acquire a common vacuum expectation value of order 
$\xi$. Our basic requirement is to choose the X-charges and the 
superpotential  so as to forbid all solutions to the vacuum 
equations except those corresponding to 
$<\theta_1,\theta_2,\cdots,\theta_m>$.

Since we work in the context of global supersymmetry  unbroken at the 
scale $\xi$, directions in the scalar field parameter space will be 
determined by the conditions $D_X=0$ and $F_i \equiv \partial W / 
\partial \Phi_i =0$. For instance, the assumption of $D_X$-flatness 
($D_X=0$ in (\ref{eq:DX})) automatically takes care of  the directions 
$<\Phi_1, \Phi_2,\cdots,\Phi_n>$ where $X_i <0$ for $i \in 
\{1,\cdots,n\}$. 

There is necessarily some gauge symmetry other than the anomalous 
$U(1)_X$, for
example the symmetries of the standard model. $D$-flatness
for these symmetries plays an important role for $S$ invariants: it tends to align the
fields present in $S$. Take for example $S=\Phi_1 \Phi_2$. Invariance
under $U(1)_Y$ implies that the hypercharges of $\Phi_1$ and $\Phi_2$ are
opposite: $Y_1 = - Y_2$. Then the corresponding $D$-term reads:
$D_Y \sim Y_1 (|\Phi_1|^2 - |\Phi_2|^2)+ \cdots$ And $D_Y = 0$
implies $<\Phi_1> = <\Phi_2> \equiv v^{}_S$. 
The contribution to $D_X$ from these fields is $x_S\vert 
v_S^{}\vert^2$, where $x_S$ is the total $X$-charge of $S$. Hence a 
positive $x_S$ will allow a vacuum with the flat direction 
$<\Phi_1\Phi_2>$.  

On more general grounds, it has been shown~\cite{invariant} that there is 
a systematic classification of the solutions to vanishing $D$-terms using 
holomorphic invariant polynomials. We will assume in what follows that 
there are enough gauge symmetries to align the fields in each of the 
invariants $S$ that we consider.     We now proceed with different 
examples with an increasing number of fields. 
\vskip .5cm 
\noindent {\bf Models with 2 fields.} 
\vskip .5cm 
\noindent {\em Model with $\theta$ 
($X$-charge $x$) and $\nb$ ($X$-charge $x_{\nb}$).}  With  these fields 
the $U(1)_X$ $D$-term is: 
\begin{equation} 
D_X = x |\theta|^2 + x_{\nb} |\nb|^2 - \xi^2\ . 
\end{equation} 
There are three different flat 
directions to consider: the desired $<\theta>$, and  $<\nb>$, and 
$<\theta,\nb>$ which we wish to avoid. The direction $<\theta>$ is 
favored by choosing $x>0$, and $<\nb>$ is forbidden if $x_{\nb}<0$. The 
third direction is allowed by $D_X^{}=0$. However since $xx_{\nb}<0$, we 
can form a holomorphic invariant involving $\nb$ and $\theta$. 

The simplest possible invariant in the superpotential is $\nb \theta^n$ 
but the corresponding $F$-terms forbid $<\theta>$ and $<\theta, \nb>$. 
Thus we lose the possibility of a $D_X$-flat direction with $<\theta> 
\sim \xi$. Since all possible flat directions are lifted, supersymmetry 
is spontaneously broken.

We must therefore require the presence of an invariant $\nb^p \theta^n$ with $p
\ge 2$ and $n \not= 0 \; mod(p)$ to forbid only the direction $<\nb,\theta>$.
The case $p=2$ corresponds precisely to a Majorana mass term for the right-handed
neutrino $\nb$, once $\theta$ is allowed a vev.
In this case, $n= 2k+1$ and there must be the following relation between 
the X-charges:
\beq
{x_{\nb} \over x } = -{ 2k+1 \over 2}\ ,
\eeq
so that $\nb$ is like a spinor, and $\theta$ a vector.
\vskip .5cm
\noindent{\em Model with $\theta$ ($X$-charge $x>0$) and $S$ ($X$-charge $x_S<0$).} With one field   
and one invariant, we have 
\begin{equation}
D_X = x |\theta|^2 + x_{S} |v_S^{}|^2 - \xi^2\ .
\end{equation}
As previously, $D_X$-flatness kills the direction $<S>$ and allows 
$<\theta>$. The main difference is that $S$ being a composite field, 
$S = \prod_{i = 1}^n \Phi_i^{n_i}$, 
the $F$-terms corresponding to the invariant $S^t \theta^u$ are $F_{j}= 
tn_j(\prod_{i \not= j} \Phi_i^{n_i})  \Phi_j^{n_j-1}S^{t-1} \theta^u$ and 
$F_\theta=uS^t \theta^{u-1}$; 
they therefore only forbid the direction $<S,\theta>$, even for $t=1$.

One is therefore left with a vev of order $\xi$ along 
the single direction $<\theta>$.
It is certainly encouraging that  linear terms in $S$
can appear in the superpotential.   Terms
such as ${\bf Q}_i  \overline{\bf u}_k H_u (\theta / M)^n$ are needed to implement 
 hierarchies among the Yukawa couplings. Conversely,  
requiring that 
\beq
{x_S\over x}=-n\ ,
\eeq
with $n$ integer $\ne 0$, is sufficient to insure the linear appearance of the 
invariant $S$. In this case, the vacuum structure is inexorably related 
to the Yukawa hierarchies. However if $x_S=0$, there is no danger associated with 
$S$, and the above discussion does not apply.  
\vskip .5cm
{\bf Models with 3 fields.}
\vskip .5cm
\noindent {\em Model with  $\theta$ ($x >0$), $\bar N$ ($x_\nb$)
and $S$ ($x_S$).} Its analysis depends on the sign of the $X$-charges of $\nb$ and $S$. 
\vskip .2cm
 A-) Let us first discuss the case $x_\nb,x_S<0$. The vanishing of the $D_X$ term 
\beq
D_X = x |\theta|^2 + x_\nb |\nb|^2 + x_S |v_S|^2 - \xi^2
\eeq
forbids the  directions $<\nb>$, $<S>$ and $<\nb,S>$, but allows the 
directions $<\theta,\nb>$, $<\theta,S>$ and $<\theta,\nb,S>$. We saw 
earlier that an invariant $\nb \theta$ forbids the desired direction 
$<\theta>$. We must require the presence of an invariant $\nb^q \theta^n$ 
with $q \ge 2$ and $n \not= 0 \; mod(q)$, to disallow the directions 
$<\theta,\nb>$ and $<\theta,\nb,S>$ ($p=2$  generates masses for $\nb$). 
The last direction $<S,\theta>$ is disposed of by adding an invariant of 
the form $S^t \theta^u$, which is also allowed given the signs of the 
charges. Consider this model for two different choices of invariants.

\vskip .2cm
 a-) Suppose $S=LH_u$ and search for charges which 
allow for the couplings (\ref{eq:Majorana}) and (\ref{eq:Dirac}) of
the seesaw mechanism. One can easily show that necessarily:
\beq
{x_\nb \over x} = -{2k+1 \over 2}\ ,\qquad {x_{LH_u} \over x} = -{2k'+1 \over 2}
\eeq
where $k$ and $k'$ are integers. This assignment automatically forbids a term 
$S \theta^q$ which would break $R$-parity. One obtains
\beq
W \ni \nb^2 \theta^{2k+1}\ , \nb LH_u \theta^{k+k'+1}\ , (LH_u)^2 
\theta^{2k'+1}\ , \cdots
\label{eq:ex1} \eeq 
where the last term does not break $R$-parity and gives an
extra contribution to neutrino masses.
\vskip .2cm
 b-) Take $S=H_dH_u$, the so-called $\mu$-term,  the mass term 
which plays a central role in all supersymmetric extensions of the 
Standard Model. Phenomenology requires in the low energy theory a term 
linear in $S$ with a mass of the order of the electroweak scale. 
If this term comes from  a 
supersymmetric term in the superpotential, one is left with two possibilities:
\begin{itemize}
\item the invariant $H_dH_u\theta^p$ which disposes of the unwanted 
direction $<S,\theta>$. The corresponding $\mu$-parameter is 
\beq
\mu \sim M \left( {\theta \over M} \right)^p ,
\eeq
as long as $x_S/x=-p$, and  phenomenology 
requires $p$ to be an extremely large integer.
\item the only other invariant linear in $S$ which kills the direction $<S,\theta>$
is $\nb S \theta^p$. One recovers the invariants of the previous example: 
eq. (\ref{eq:ex1}) or more generally if we assume a coupling $\nb^q \theta^{qk+r}$
($0<r<q$ and $q \ge 2$):
\beq 
\nb^q \theta^{qk+r},\; \; \nb H_dH_u \theta^{k+k'+1},\; \; (H_dH_u)^q \theta^{q(k'+1)-r}
\eeq
where $k+k'+1=p$.
The $\mu$-parameter reads:
\beq
\mu = <\nb> \left( {\theta \over M} \right)^{k+k'+1}.
\eeq
This is somewhat a more hopeful situation since we expect in any case that 
$<\nb> \ll M$ and $<\theta>/M \ll 1$. But it requires $<\nb> \not= 0$ and
the coupling $\nb S \theta^p$ breaks $R$-invariance if $\nb$ is to be interpreted 
as a right-handed neutrino ({\em i.e.} if it has a non-zero Dirac-type coupling).
If $q=2$, then $\nb$ mixes with the right-handed neutrinos and has $R$-parity 
equal to $-1$.

Alternatively if there is no term in the superpotential linear in 
$H_dH_u$, there may be a non-holomorphic one in the K\"ahler potential~\cite{GM}. 
The only possible choice is
\beq
K \ni H_d H_u \nb^{*p} + \; {\rm h.c.}\ ,
\eeq
but in this case,
\beq
\mu = m_{3/2} \left( {<\nb^*> \over M} \right)^p,
\eeq
and since we know that  $<\nb> \ll M$, we obtain $\mu \ll m_{3/2}$ which 
is probably too small for phenomenological applications.
\end{itemize}
\vskip .2cm

 B-) We now turn to the case $x_\nb<0$ and $x_S>0$. The vanishing 
of  $D_X$ avoids only $<\nb>$.
The need to dispose of the $<S>$ direction imposes an invariant $\nb S$ 
(hence $x_\nb=-x_S$)
which forbids the directions $<\nb>$, $<S>$, $<\nb,S>$, and $<\nb,S,\theta>$. The last 
direction $<\nb,\theta>$ is taken care of by an invariant $\nb^q 
\theta^n$ ($n\not= 0 \; mod(q)$),
which in the simplest case is $\nb^2 \theta^{2k+1}$.
\vskip .2cm
a-) When $S=LH_u$, the Dirac neutrino mass term ($\nb S$) appears here with no 
suppression factor $<\theta>/M$.
\vskip .2cm
 b-) When  $S$ is the $\mu$-term, it cannot appear by itself in 
the superpotential, although it is allowed in the K\"ahler 
potential~\cite{GM} as
\beq
K \ni H_d H_u \theta^{*q} + \; {\rm h.c.}\ .
\eeq
But since the $X$-charge of $H_dH_u$ must be half-odd integer (to forbid 
the invariant 
$\nb \theta^m$ that would suppress the direction $<\theta>$), this term 
cannot appear in the K\"ahler potential either. 
Hence the presence of a singlet $\nb$ with charge $x_\nb<0$ is enough to 
forbid this interesting possibility for generating a
low-energy mu-term~\cite{Nir}.
One can still generate a $\mu$-term through the 
invariant $\nb H_dH_u$ in the superpotential, with 
\beq
\mu \sim \; <\nb>.
\eeq
\vskip .5cm
\noindent {\em Model with  one field $\theta$ ($x >0$), and two composites, $S_1$ ($x_1$)
and $S_2$ ($x_2$).}
\vskip .3cm
If one of the $S$ has a positive charge (say $x_1>0$),   we are left with the 
corresponding direction ($<S_1>$) since no $F$-term can kill a single $<S>$ and the 
$D_X$-term only forbids the other $<S>$ ($<S_2>$).

We therefore require both $x_1,x_2 <0$, in which case $D_X$-flatness deals with $<~S_1>$, $<S_2>$ 
and $<S_1,S_2>$, the invariant $S_1^t \theta^u$ with $<S_1,\theta>$ and $<~S_1,S_2,\theta>$,
and the invariant $S_2^r \theta^v$ with the remaining $<S_2,\theta>$. 
When $|x_1|/x$ and $|x_2|/x$ are integers, this can be used to generate mass 
hierarchies in the quark sector~\cite{IR}-\cite{PMR} with:
\beq
S_1 = {\bf Q}  \overline{\bf d}H_d\ , \; \; \; S_2 = {\bf Q}  
\overline{\bf u}H_u\ .
\eeq
We come to the important conclusion that the mere determination of the 
$\xi$-vacuum requires  Yukawa terms 
in the superpotential.  Alternatively, their absence from the 
superpotential (supersymmetric zeros)  can unleash unwanted flat
directions in the $\xi$-vacuum.  
\vskip .5cm
\noindent{\em Models with  $\theta$ ($x >0$), $\nb_1$ ($x_1$)
and $\nb_2$ ($x_2$).}
\vskip .3cm 

 If $x_1,x_2<0$, $D_X$-flatness excludes $<\nb_1>$, $<\nb_2>$ and $<\nb_1,\nb_2>$. 
Requiring Majorana mass terms imposes as seen above that $|x_1|/x = (2k+1)/2$
and $|x_2|/x = (2l+1)/2$. The Majorana mass matrix then reads:
\beq
{\cal M}_M \sim \theta \left( \begin{array}{cc} \theta^{2k} & \theta^{k+l} \\
                                         \theta^{k+l} & \theta^{2l} \end{array} \right) .
\eeq

 If $x_1<0$ and $x_2>0$, only the direction $<\nb_1>$ is disposed of by $D_X$. We need an
invariant of the form $\nb_1 \nb_2^l$ to kill the direction $<\nb_2>$  and an invariant 
$\nb^2_1 \theta^{2k+1}$ to deal with $<\nb_1, \theta>$. This in turn imposes constraints 
on the $X$-charges:
\beq
{x_1 \over x} = -{2k+1 \over 2},\; \; \; {x_2 \over x} = {2k+1 \over 2l},
\eeq
If $l=m(2k+1)$, the lowest invariants in the superpotential are
\beq
W \ni \nb_1^2 \theta^{2k+1},\; \nb_1 \nb_2^{(2k+1)m},\; \nb_1 \nb_2^m \theta^k.
\eeq
These three invariants obey a polynomial constraint  and 
do not suffice to eliminate $<\nb_2,\theta>$, along 
which $F_{\nb_1}$ can vanish, as two invariants are linear in $\nb_1$. 

A case of interest is $m=1$ for which one obtains a mixed Majorana
mass term $\nb_1 \nb_2 \theta^k$. For $m\ne 1$, one of the $\nb$ stays 
massless after $\xi$-breaking.
\section{Mu-term}

We will not continue a general study adding fields one by one. The 
analysis becomes more and more involved as the number of fields increases 
but one can always restrict to a subset of fields and the previous 
examples with two or three fields may provide good insights to treat the 
more complicated cases. Let us illustrate this on an example which makes 
heavy use of models a-) and b-) discussed above, in which additional 
constraints on the charges emerge naturally. 

It is well-known that the gauge symmetries of the supersymmetric 
standard model do not make any distinction  between the invariants $S_0 
\equiv H_d H_u$ and $S_i \equiv L_i H_u$ ($i$ being a family index). The 
first one appears in the $\mu$-term whereas the others correspond to 
R-parity violating terms in the superpotential. As discussed in model 
a-), they also appear in the seesaw mechanism in conjunction with a 
Standard Model singlet $\nb_i$ which plays the role of a right-handed 
neutrino. Similarly, the discussion of model b-) leads to the conclusion 
that a possible solution to the $\mu$-problem involves another 
standard model singlet $\nb_0$. We therefore consider the following set 
of fields and $X$-charges
\beq
\begin{array}{cccccc}
      & \theta & \nb_0 & S_0 & \nb_i & S_i \\
X/x=\;  & 1 & -{3k_0 +1\over 3} & -{3k'_0 +2\over 3} & 
-{2k_i +1\over 2} & -{2k'_i +1\over 2} \end{array}
\label{eq:(M+1)SSM} \eeq 
The superpotential then includes the terms
\beq
W \ni \nb_i S_j \theta^{k_i + k'_j +1}, \; \nb_i \nb_j \theta^{k_i + k_j +1}, \;
\nb_0 S_0 \theta^{k_0+k'_0 +1},\; \nb_0^3 \theta^{3k_0+1}, \cdots
\label{eq:W} \eeq
(terms mixing $\nb_0$ and $\nb_i$ are highly non-renormalisable).
It is invariant under a $R_p$ parity defined as $+1$ for $\nb_0$ and $S_0$ and 
$-1$ for $\nb_i$ and $S_i$. This is why we have not chosen half-odd charges 
for $\nb_0$ (and thus $S_0$): mixed terms $\nb_0 \nb_i$ would have led to $R_p$
violations.

There are no supersymmetric mass term for $\nb_0$; it is therefore induced 
by supersymmetry breaking and it is of the order of the scale of supersymmetry
breaking $\tilde m$. Assuming that renormalisation group evolution turns
the corresponding scalar  mass-squared term negative, the potential for the 
$\nb_0$ scalar field has the form:
\beq
V = - \tilde m^2 |\nb_0|^2 + {\lambda \over 2} |\nb_0|^4 \left( 
{\theta \over M} \right)^{2(3k_0+1)}
\eeq
Its minimization leads to the following value for the $\mu$-term:
\beq
\mu \sim {\tilde m \over \sqrt{\lambda}} \left( {\theta \over M} \right)^{k'_0 - 2 k_0}.
\eeq
The terms involving $\nb_i$ and $S_i$ in (\ref{eq:W}) are precisely the ones 
necessary to implement the seesaw mechanism. 

Let us note here that if we had taken the more general charges for $\nb_0$ and
$S_0$: $${X_{S_{{0}}} \over x} = - {(2p+1)k'_0+r_0 \over 2p+1}, \; \; \;
{X_{\nb_{{0}}} \over x} = - {(2p+1)(k_0+1)-r_0 \over 2p+1}$$
with $p \ge 1$ and $0 < r_0 < 2p+1$,  we would have obtained a mu-term of order:
\beq
\mu \sim \left( {\tilde m M^{2(p-1)} \over \sqrt{\lambda}} \right)^{1/(2p-1)} 
\left( {\theta \over M} \right)^{[k'_0(2p-1) -2k_0 +r_0-2]/(2p-1)}.
\eeq
If we take $M$ to be of the order of the Planck scale, the first factor is a scale 
intermediate between $\tilde m$ and $M_{Pl}$ (unless $p=1$) and the suppression 
due to the second factor can only be minor. We  conclude that we must take
$p=1$ in order to have a low energy mu-term.  We are thus led to the choice 
(\ref{eq:(M+1)SSM}) of quantum numbers. The corresponding model with the superpotential 
terms (\ref{eq:W}) leads at low energy to the so-called (M+1)SSM model (minimal 
supersymmetric model plus an extra singlet), with the 
important difference that the field $\nb_0$ is not a complete gauge singlet: it is
charged at least under the anomalous $U(1)_X$ symmetry.

We discussed in detail the previous case because it is illustrative of 
the power of the method and how far one can take it. Let us now summarize 
for further use the different scenarios that we have encountered, 
depending on the value of the charge 
\beq
X^{[\mu]} \equiv X_{H_{u}} + X_{H_{d}}
\eeq
and possibly the charge of the associated singlet $X_{\nb_{0}}$. We only consider scenarios 
which do not lead to an obvious breaking of $R$-parity (for example if we need $<\nb_0> \not= 0$, 
then $\nb_0$ has $R$-parity $+1$). 
\begin{itemize}
\item if $X^{[\mu]}/x = -p_0$ with $p_0$ integer, then
\beq
\mu = M \left( {\theta \over M} \right)^{p_0}
\eeq
\item if $X^{[\mu]}/x = -(3k'_0 + r_0)/3$ and 
$X_{\nb_{0}} /x = - (3k_0 + 3 - r_0)/3$ with $k_0$ and $k'_0$ 
integers and $r_0=1$ or $2$, then
\beq
\mu = <\nb_0> \left( {\theta \over M} \right)^{k_0+k'_0+1}.
\eeq
This is the case just discussed.
\item if $X^{[\mu]}/x = -X_{\nb_{0}} / x = (3k_0 + r_0)/3$
with $k_0$ integer and $r_0=1$ or $2$, then
\beq
\mu = <\nb_0>.
\eeq
\item finally, if $X^{[\mu]} = 0$, the $H_dH_u$ term is unseen by the anomalous symmetry
and therefore not likely to lead to vevs of order $\xi$ for the Higgs doublets, 
in the direction where the $U(1)_X$ is broken: indeed the direction $<H_u,H_d>$ is 
taken care of by the requirement of $D_X$-flatness. If the $\mu$-term 
does not appear in the superpotential, supersymmetry breaking can then 
lift the remaining flat direction ($|<\theta>| = \xi/\sqrt{x}$, $|<H_u>|=|<H_d>| =v$)~\cite{DGPS}. 
\end{itemize}
We will see below that anomaly 
cancellation conditions tend to give integer values for $X^{[\mu]}$, which disfavours the 
second and third possibilities. And the first one requires too large values of $p_0$
if $M$ is of the order of the Planck scale. The last solution 
($X^{[\mu]}=0$), together with the absence of the 
$\mu$-term in the superpotential, seems to us favoured. We offer no 
reason for this absence, except by deferring to string lore according to 
which there are no mass terms in the superpotential.

\section{R-parity}

We now show that the constraints discussed above on the $U(1)_X$ 
quantum numbers of the low-energy fields may naturally lead to
conserved $R$-parity. The presence of standard model singlets $\nb_i$
necessary to implement the seesaw mechanism plays in this respect a key role.

We assume the seesaw mechanism requiring the presence of the invariants:
\beq
\bar N_i \bar N_j \theta^{n^0_{ij}}
        + L_i \bar N_j H_u \theta^{n^{\nu}_{ij}} 
\label{eq:seesaw}
\eeq
We saw that, in order not to spoil the $\xi$-vacuum, the powers $n^0_{ii}$ must
be odd integers, or equivalently the $X$-charges $X_{\nb_{i}}$ of the fields
$\bar N_i$ must be, in units of $x$, half-odd integers:
\beq
{X_{\nb_{i}} \over x} = - \frac{2k_i+1}{2}
\eeq
where $k_i$ is an integer. Henceforth we set $x=1$. The last term in (\ref{eq:seesaw}) 
determines the $R$-parity of the right-handed neutrino superfields to be negative.

Let us study the $X$-charges of possible standard model
invariant operators made up of the basic fields $ {\bf Q}_i$, $\bar {\bf 
u}_i$, $\bar {\bf d}_i$, $L_i$, $\bar e_i$ ($i$ being  a family index) and of the
Higgs fields $H_u$, and $H_d$.

The cubic standard model invariants that respect baryon and lepton
numbers are, in presence of the gauge singlets $\nb_i$,
\beq
{\bf Q}_i \bar {\bf d}_j H_d\ ,~~ {\bf Q}_i\bar  {\bf u}_j H_u\  ,~~ L_i \bar e_j H_d\ ,~~ L_i \nb_j H_u\ ,
\eeq
with charges $X^{[d]}_{ij}$, $X^{[u]}_{ij}$, $X^{[e]}_{ij}$ and
$X^{[\nu]}_{ij}$ respectively.
To avoid undesirable flat directions,  all must appear in the
superpotential, restricting   their $X$-charges to be of the form
\beq
X^{[u,d,e,\nu]}_{ij}=-n^{u,d,e,\nu}_{ij} \ ,
\eeq
where $n^{u,d,e,\nu}_{ij}$ are all positive integers or zero.

We now turn to the invariants which break $R$-parity. We have already 
encountered the quadratic invariants $L_i H_u$ whose charges are determined
by the seesaw couplings (\ref{eq:seesaw}) to be half odd integers
\beq
X_{L_{i} H_u}={2\ (k_j-n^\nu_{ij})+1\over 2}\ \ .
\eeq
Consider the cubic $R$-parity violating operators,
$L_i L_j \bar e_k$, $L_i {\bf Q}_j \bar {\bf d}_k$ and $\bar {\bf u}_i 
\bar {\bf d}_j \bar {\bf d}_k$.
The charges of the first two, which violate lepton number,
satisfy the relations:
\beq
X_{L_i L_j \bar e_k} = X^{[e]}_{jk} + X^{[\nu]}_{il} - X^{[\mu]} - X_{\nb_{l}}
\eeq
\beq
X_{L_i Q_j \bar d_k} = X^{[d]}_{jk} + X^{[\nu]}_{il} - X^{[\mu]} - X_{\nb_{l}}
\eeq
where the index $l$ can be chosen arbitrarily. 

As a consequence, if $X^{[\mu]}$ is integer, $ -p_0$, both charges
are half-odd integers and there is no $R$-parity violation from these 
operators. However they  can still appear  as
\bea
L_i L_j \bar e_k \bar N_l \ \theta^{(n^e_{jk}+n^{\nu}_{il}-p_0)} & &
L_i Q_j \bar d_k \bar N_l \ \theta^{(n^d_{jk}+n^{\nu}_{il}-p_0)}
\eea
in the superpotential. A similar conclusion is reached if $X^{[\mu]}$ is
a multiple of one third, in which case one needs to include also apropriate powers of 
$\nb_0$.

To determine the charges of the operators $\bar {\bf u}_i \bar 
{\bf d}_j \bar {\bf d}_k$ in terms of the charges of 
the parity-conserving invariants, one must use
the Green-Schwarz  condition on mixed anomalies $C_{\rm weak}=C_{\rm color}$ which reads:
\beq
\sum_i (X_{Q_{i}} + X_{L_{i}}) - \sum_i (X_{\bar u_{i}} + X_{\bar d_{i}}) + X^{[\mu]} = 0.
\label{eq:C2C3}
\eeq
One obtains:
\bea
X_{\bar u_{i} \bar d_{j} \bar d_{k}} &=&
{1 \over n^2_f} \sum_{p,q} X^{[d]}_{pq}
+ {1 \over n_f} \sum_l \left[ X^{[u]}_{pi} + X^{[d]}_{pj}
+ X^{[d]}_{pk} - X^{[u]}_{pl} - 2 X^{[d]}_{pl} + X^{[\nu]}_{lm} \right] \nonumber \\
& & + {1-n_f \over n_f} X^{[\mu]} - X_{\nb_{_m}}\ ,
\label{eq:udd}
\eea
true for any two family indices $p,m$, and  where $n_f$ is the number of 
families which we will take to be three. In a large class of models, the 
charge $X_{\bar u_{i} \bar d_{j} \bar d_{k}}$ thus obtained will be such 
as to forbid not only a term $\bar u_{i} \bar d_{j} \bar d_{k}$ in the 
superpotential but also any term obtained from it by multiplying by any 
powers of $\theta$, $\nb_i$ or $\nb_0$.

Let us consider for illustrative purpose
an anomalous symmetry which is family independent. Then
(\ref{eq:udd}) simplifies to:
\beq
X_{\bar {\bf u} \bar 
{\bf d} \bar {\bf d}}=X^{[d]}+X^{[\nu]}
-(1-{1\over n_f})X^{[\mu]}-X_{\nb}\ .
\label{eq:famind}
\eeq
Remember that $X_{\nb}$ is half-odd integer and $n_f = 3$. If $X^{[\mu]}$ 
is integer not proportional to $n_f=3$, then the charge $X_{\bar {\bf u} 
\bar {\bf d} \bar {\bf d}}$ is such that no term $\bar {\bf u} \bar {\bf 
d} \bar {\bf d} \theta^m \nb^n$ can be invariant.  If $X^{[\mu]}$ is 
non-integer and a multiple of one third, then similarly no term $\bar 
{\bf u} \bar {\bf d} \bar {\bf d}\theta^m \nb^n \nb_0^p$ can be made 
invariant.  In the low energy theory,  baryon number violation becomes 
negligible. 

If we restrict our attention to models which yield $\sin^2 \theta_W = 3/8$, 
the Green-Schwarz condition $5C_{\rm weak} = 3 C_Y$ reads:
\beq
\sum_i (7X_{Q_{i}} + X_{L_{i}}) - \sum_i (4X_{\bar u_{i}} + X_{\bar d_{i}}
+ 3 X_{\bar e_{i}}) + X^{[\mu]} = 0 .
\label{eq:C2C1}
\eeq
One infers from (\ref{eq:C2C3}) and (\ref{eq:C2C1}) the following relation:
\beq
X^{[\mu]} = \sum_i \left( X^{[d]}_{ii} - X^{[e]}_{ii} \right),
\eeq
which tends to favor models with integer $X^{[\mu]}$ (proportional to $n_f$ in the 
case of a family-independent symmetry).

If $X^{[\mu]} = 0$ or more generally if $X^{[\mu]}$ is proportional to $n_f$ 
($X^{[\mu]} = n_f z_{\mu}$), the charge in (\ref{eq:famind}) is half-odd integer; 
it can only be
compensated by odd powers of $\bar N$: invariance under $X$ means conservation
of $R$-parity. For instance, the above allows the interaction:
\beq
\bar{\bf u}\bar{\bf d}\bar{\bf d}\bar N~\theta^{[n_d+n_\nu+z_\mu(n_f-1)]}_{}\ .
\eeq
This term allows baryon number violation, but preserves both $B-L$ and $R$-parity.

A very similar discussion can obviously be given for the general case of a 
family dependent anomalous symmetry.

To conclude, in a large class of models, there are no $R$-parity violating operators,
whatever their dimensions: through the right-handed
neutrinos for example, $R$-parity is linked to half-odd integer charges, so that
$U(1)_X$ charge invariance results in $R$-parity invariance. Thus   none of
the operators that violate $R$-parity can appear in holomorphic invariants:
even after breaking of the anomalous $X$ symmetry, the remaining
interactions all respect $R$-parity, leading to an  absolutely
stable superpartner.

\section{Conclusion}

The purpose of this work was to show that the breaking of an anomalous 
Abelian gauge symmetry  at a very large scale imposes some very stringent 
constraints on the anomalous charges of the low energy fields. This is 
indeed the situation in many superstring models (see for example 
refs.~\cite{faraggi,orbifold}).   Holomorphy and $D$-flatness, especially 
for the anomalous symmetry, are the tools that we used to derive these 
constraints. This in turn puts some restrictions on the dynamics of the 
model. We illustrated this fact on the generation of a mu-term. If we 
consider a model with a single extra $U(1)$ and a single field ($\theta$) 
which acquires a {\em vev} when the anomalous symmetry is broken, it 
tends to exclude most of the scenarios proposed to generate a mu-term. 
The preferred scenario seems to be a mu-term invariant under the 
anomalous symmetry. It is an open question why such a term does not 
appear in the superpotential but rather in the K\"ahler potential. In our 
discussion, we tried to impose conditions which lift most of the 
dangerous flat directions. Some of them might however require 
supersymmetry breaking and it remains to be seen whether, there also, the 
anomalous $U(1)$  symmetry plays a role~\cite{BD,DP}. 

The most interesting result is that, once these constraints are taken 
into account, the couplings which respect the anomalous symmetry also 
respect $R$-parity. This remains true at low energy since we made the 
supposition that the field $\theta$ which breaks this symmetry has 
$R$-parity $+1$. We derived this result in the restricted class of models 
discussed in this paper but we believe that it is rather general. It 
makes use of the  right-handed neutrinos necessary to generate neutrino 
masses through the seesaw mechanism. This should lead to an interesting 
phenomenology of lepton number or baryon number violations.

Of course, although we tried to be more general in this paper, it is 
tempting to apply these constraints to models of quark and lepton mass 
hierarchies. It was found~\cite{BR} that, among models using an Abelian 
gauge symmetry, the observed hierarchies favor a model with non-zero 
mixed anomalies satisfying the relation $5C_{weak} = 3 C_1$ (if we assume 
$C_{weak} =  C_{color}$ and the charge of the anomalous mu-term to be 
zero~\cite{Nir}). From the discussion of the last section, this seems in 
complete agreement with what is required by the constraints discussed in 
this paper. Of course, if we want to obtain a realistic quark and lepton 
mass spectrum, we need to introduce several extra $U(1)$ (a single 
combination of which is anomalous) and several $\theta$ 
fields~\cite{PMR}. The discussion then becomes more involved but the 
simple case discussed in the present work leads us to expect possible 
rewards such as $R$-parity. And it is a fact, often neglected, that any 
theory of mass has to come up with solutions for $R$-parity or to explain 
why its violations are mild.  
\vskip 1cm 
{\bf Acknowledgments:} 
\vskip .5cm 
P. B. \& R. wish to thank the Aspen Center for Physics where 
this work was started. S. L. thanks the Institute for Fundamental Theory, 
Gainesville, for its hospitality and financial support. Part of this work 
was supported by a CNRS-NSF grant INT-9512897.

\end{document}